\documentclass[12pt]{article}
\usepackage{a4}
\usepackage{amsfonts}
\usepackage{amsmath}
\usepackage{latexsym}
\usepackage{amsfonts}

\begin{document}

\renewcommand{\theequation}{\arabic{section}.\arabic{equation}}
\thispagestyle{empty}
\vspace*{-1,5cm}
\noindent \vskip3.3cm

\begin{center}
{\Large\bf Off-shell construction of some trilinear higher spin gauge field interactions }

{\large Ruben Manvelyan ${}^{\dag\ddag}$, Karapet Mkrtchyan${}^{\ddag}$ \\and Werner R\"uhl
${}^{\dag}$}
\medskip

${}^{\dag}${\small\it Department of Physics\\ Erwin Schr\"odinger Stra\ss e \\
Technical University of Kaiserslautern, Postfach 3049}\\
{\small\it 67653
Kaiserslautern, Germany}\\
\medskip
${}^{\ddag}${\small\it Yerevan Physics Institute\\ Alikhanian Br.
Str.
2, 0036 Yerevan, Armenia}\\
\medskip
{\small\tt manvel,ruehl@physik.uni-kl.de; karapet@yerphi.am}
\end{center}\vspace{2cm}

\bigskip
\begin{center}
{\sc Abstract}
\end{center}
\quad Several trilinear interactions of higher spin fields involving two equal ($s=s_{1}=s_{2}$) and one higher even ($s_{3}\geq 2s$) spin are presented. Interactions are constructed on the Lagrangian level using Noether's procedure together with the corresponding next to free level fields of the gauge transformations. In certain cases when the number of derivatives in the transformation is $2s-1$ the interactions lead to the currents constructed from the generalization of the gravitational Bell-Robinson tensors. In other cases when the number of derivatives in the transformation is more than $2s-1$ we obtain the finite tower of interactions with smaller even spins less than $s_{3}$ in full agreement with our previous results for the interaction of the higher even spins field with a conformal scalar \cite{MMR, MR}.

\newpage

\section{Introduction}
\quad The construction of interacting \emph{higher spin gauge field} theories (HSF) has always been considered an important task during the last thirty years (See \cite{vanDam}-\cite{Petkou} and ref. there\footnote{We do not pretend here for complete quotations and just present some references important for us during this investigation}).
The complications and difficulties which accompany any serious attempt to solve the essential problems in this area always attracted interest but activity intensified after discovering the important role HSF plays in $AdS/CFT$ correspondence. Particular attention caused the holographic duality between the $O(N)$ sigma model in three dimensional space and HSF gauge theory living in the four dimensional space with negative constant curvature \cite{Kleb}. This case of holography is singled out by the existence of two conformal points of the
boundary theory and the possibility to describe them by the
same HSF gauge theory with the help of spontaneously breaking of higher spin gauge symmetry and mass generation by a corresponding Higgs mechanism. All these complicated physical tasks necessitate \emph{quantum loop} calculations for HSF field theory \cite{MMR1}-\cite{MR5} and therefore information about manifest, off-shell and Lagrangian formulation of possible interactions for HSF. Then after successful calculations on the quantum level the construction can be controlled by comparison with the boundary $O(N)$ model results checking the $AdS/CFT$ correspondence conjecture on the loop level \cite{MMR1}, \cite{Ruehl}, \cite{MR2}.

In this article we continue the construction of possible couplings including different higher spin fields which was started in our previous articles about couplings including HSF and scalar fields \cite{MMR, MR, MMR1} and that are important for the Higgs mechanism mentioned above. Here we turn to the trilinear interaction between HSF gauge fields of different spins (s-s-s') in a flat background but the results can be easily generalized to the $AdS$ background. The first three sections are devoted to the development of the idea: how we can apply higher spin gauge symmetry of a spin "s" gauge field to the field with a spin lower than "s". Then getting in this way information about the first order gauge transformation, we can handle Noether's procedure applying this first order transformation to the zero order free Lagrangian and integrating this variation to a first order trilinear interaction.
Starting from the construction of the spins 1-1-2 and 1-1-4, we discover in this simple case the same phenomenon as in the previously investigated scalar case \cite{MMR, MR}, namely the appearance of the couplings with all even spins lower than the initial maximal higher spin gauge field  involved in the interaction vertex
(Section 2). Then we generalize the construction to the more complicated 2-2-4 and then to 2-2-6 where the previously constructed 2-2-4 interaction again appears automatically (Section 3). The next section starts from the description of a technique for working with the HSF fields in Fronsdal's  \cite{Frons} formulation and deWit-Friedman curvatures \cite{DF, MR6}. In the same section we succeed with the construction of the interaction Lagrangian of spin type s-s-2s together with the first order higher spin gauge transformation.

\section{Exercises on spin one field couplings with the higher spin gauge fields}
\quad We start this section constructing the well known interaction of the  electromagnetic field $A_{\mu}$ in flat $D$ dimensional space-time with the linearized spin two field. Hereby we illustrate how Noether's  procedure regulates the relation between gauge symmetries of different spin fields. The standard free Lagrangian of the electromagnetic field is
\begin{eqnarray}
&&\mathcal{L}_{0}=-\frac{1}{4}F_{\mu\nu}F^{\mu\nu}=-\frac{1}{2}\partial_{\mu}A_{\nu}\partial^{\mu}A^{\nu}+\frac{1}{2}(\partial A)^{2} ,\label{L0}\\
&&F_{\mu\nu}=\partial_{\mu}A_{\nu}-\partial_{\nu}A_{\mu} ,\quad \partial A=\partial_{\mu}A^{\mu} .
\end{eqnarray}
To construct the interaction we propose a possible form for the action of the spin two linearized gauge symmetry
\begin{equation}\label{gi}
    \delta_{\varepsilon}^{0}h^{(2)\mu\nu}(x)=2\partial^{(\mu}\varepsilon^{\nu)}(x)=\partial^{\mu}\varepsilon^{\nu}(x)
    +\partial^{\nu}\varepsilon^{\mu}(x) ,
\end{equation}
on the spin one gauge field $A_{\mu}(x)$. Then Noether's procedure fixes this coupling (1-1-2 interaction) of the electromagnetic field with linearized gravity correcting when necessary the proposed transformation.

We  start from the following general ansatz for a gauge variation of $A_{\mu}$ with respect to a spin 2 gauge transformation with vector parameter $\varepsilon^{\rho}$
\begin{eqnarray}
\delta_{\varepsilon}^{1}A_{\mu}=-\varepsilon^{\rho}\partial_{\rho}A_{\mu}+C\varepsilon^{\rho}\partial_{\mu}A_{\rho} \label{delta2A}.
\end{eqnarray}
Then we apply this variation (\ref{delta2A}) to (\ref{L0}) and after some algebra neglecting total derivatives we obtain
\footnote{From now on we will never make a difference between a variation of the Lagrangians or the actions discarding all total derivative terms and admitting partial integration if necessary. For compactness  we introduce also shortened notations for divergences of the tensorial symmetry parameters
\begin{eqnarray}
\epsilon_{(1)}^{\mu\nu\dots}=\nabla_{\lambda}\epsilon^{\lambda\mu\nu\dots},\quad \epsilon_{(2)}^{\mu\dots}=\nabla_{\nu}\nabla_{\lambda}\epsilon^{\nu\lambda\mu\dots}, \quad\dots
\end{eqnarray}}
\begin{eqnarray}
\delta_{\varepsilon}^{1}\mathcal{L}_{0}&=&\partial^{(\mu}\varepsilon^{\nu)}\partial_{\mu}A_{\rho}\partial_{\nu}A^{\rho}
-\frac{1}{2}\varepsilon_{(1)}\partial_{\mu}A_{\nu}\partial^{\mu}A^{\nu}
+\frac{1}{2}\varepsilon_{(1)}(\partial A)^{2}
+C\partial^{(\mu}\varepsilon^{\nu)}\partial_{\rho}A_{\mu}\partial^{\rho}A_{\nu}\nonumber\\
&-&2C\partial^{(\mu}\varepsilon^{\nu)}\partial_{\rho}A_{(\mu}\partial_{\nu)}A^{\rho}
+\frac{C}{2}\varepsilon_{(1)}\partial_{\mu}A_{\nu}\partial^{\nu}A^{\mu}
-\frac{C}{2}\varepsilon_{(1)}(\partial A)^{2}\nonumber\\
&+&(C-1)(\partial A)\partial^{\mu}\varepsilon^{\nu}\partial_{\nu}A_{\mu}.\label{delta1L0}
\end{eqnarray}
Then we have to compensate (or integrate) this variation using the gauge variation of the spin 2 field (\ref{gi}) and its trace $\delta_{\varepsilon}^{0}h_{\mu}^{(2)\mu}=2\varepsilon_{(1)}$ . We see immediately that the last line in (\ref{delta1L0}) is irrelevant but can be dropped by choice of the free  constant $C=1$. With this choice we have instead of (\ref{delta2A})
\begin{eqnarray}
\delta_{\varepsilon}^{1}A_{\mu}=-\varepsilon^{\rho}\partial_{\rho}A_{\mu}+\varepsilon^{\rho}\partial_{\mu}A_{\rho}
=\varepsilon^{\rho}F_{\mu\rho},\label{git}
\end{eqnarray}
so that our spin two transformation now is manifestly gauge invariant with respect to the spin one gauge invariance
\begin{eqnarray}
\delta^{0}_{\sigma}A_{\mu}=\partial_{\mu}\sigma,\label{sigma}
\end{eqnarray}
and our spin one gauge invariant free action (\ref{L0}) keeps this property also after spin two gauge variation. Namely (\ref{delta1L0}) now can be written as
\begin{eqnarray}
\delta_{\varepsilon}^{1}\mathcal{L}_{0}=\partial^{(\mu}\varepsilon^{\nu)}F_{\mu\rho}F_{\nu}^{\ \rho}-\frac{1}{4}\varepsilon_{(1)}F_{\mu\nu}F^{\mu\nu} .\label{211}
\end{eqnarray}
This variation can be compensated introducing the following 2-1-1 interaction
\begin{eqnarray}
\mathcal{L}_{1}(A_{\mu},h^{(2)}_{\mu\nu})=\frac{1}{2}h^{(2)\mu\nu}\Psi^{(2)}_{\mu\nu} ,\label{L1(2)}
\end{eqnarray}
where
\begin{eqnarray}
\Psi^{(2)}_{\mu\nu}=-F_{\mu\rho}F_{\nu}^{\ \rho}+\frac{1}{4}g_{\mu\nu}F_{\rho\sigma}F^{\rho\sigma} ,
\end{eqnarray}
is the well known energy-momentum tensor for the electromagnetic field.

Thus we solved Noether's equation
\begin{equation}\label{Ne}
    \delta_{\varepsilon}^{1}\mathcal{L}_{0}(A_{\mu})+\delta_{\varepsilon}^{0}\mathcal{L}_{1}(A_{\mu},h^{(2)}_{\mu\nu})=0
\end{equation}
in this approximation completely, defining a first order transformation and interaction term at the same time.
Finally note that the corrected Noether's procedure spin two transformation of the spin one field (\ref{git}) can be written as a combination of the usual reparametrization for the contravariant vector $A_{\mu}(x)$ (non invariant with respect to (\ref{sigma})) and spin one gauge transformation with the special field dependent choice of the parameter $\sigma(x)=\varepsilon^{\rho}(x)A_{\rho}(x)$
\begin{eqnarray}
\delta_{\varepsilon}^{1}A_{\mu}=\varepsilon^{\rho}F_{\mu\rho}=-\varepsilon^{\rho}\partial_{\rho}A_{\mu}
-\partial_{\mu}\varepsilon^{\rho}A_{\rho}+\partial_{\mu}\left(\varepsilon^{\rho}(x)A_{\rho}(x)\right),\label{git1}
\end{eqnarray}
A symmetry algebra of these transformations can be understood from the commutator
\begin{eqnarray}
  && [\delta_{\eta}^{1},\delta_{\varepsilon}^{1}]A_{\mu}(x)=\delta_{[\eta,\varepsilon]}^{1}A_{\mu}(x)+\partial_{\mu}
  \left(\varepsilon^{\rho}\eta^{\lambda}F_{\rho\lambda}(x)\right) \\
&& [\eta,\varepsilon]^{\lambda}=\eta^{\rho}\partial_{\rho}\varepsilon^{\lambda}-\varepsilon^{\rho}\partial_{\rho}\eta^{\lambda}
\end{eqnarray}
So we see that the algebra of transformations (\ref{git1}) close on the field dependent gauge transformation (\ref{sigma}) with parameter $\sigma(x)=\varepsilon^{\rho}\eta^{\lambda}F_{\rho\lambda}(x)$.

Now we turn to the first nontrivial case of the vector field interaction with a spin four gauge field with the following zero order spin four gauge variation
\begin{eqnarray}
\delta_{\epsilon}^{0}h^{\mu\rho\lambda\sigma}=4\partial^{(\mu}\epsilon^{\rho\lambda\sigma)},\quad
\delta_{\epsilon}^{0}h_{\rho}^{\ \rho\lambda\sigma}=2\epsilon_{(1)}^{\lambda\sigma}.\label{h4}
\end{eqnarray}
where we have a symmetric and traceless gauge parameter $\epsilon^{\mu\nu\lambda}$ to construct a gauge variation for $A_{\mu}$.
In this case we  first present final result and then explain details of the derivation.

The solution of the corresponding Noether's equation
\begin{equation}\label{Ne1}
    \delta_{\varepsilon}^{1}\mathcal{L}_{0}(A_{\mu})
    +\delta_{\varepsilon}^{0}\mathcal{L}_{1}(A_{\mu},h^{(2)}_{\mu\nu},h^{(4)}_{\mu\nu\lambda\rho})=0 ,
\end{equation}
after field redefinitions is the linearized Lagrangian for the coupling of the electromagnetic field to the spin four and spin two fields
\begin{eqnarray}
\mathcal{L}_{1}(A_{\mu},h^{(2)\mu\nu},h^{(4)\mu\nu\alpha\beta})=\frac{1}{4}h^{(4)\mu\nu\alpha\beta}\Psi^{(4)}_{\mu\nu\alpha\beta}
+\frac{1}{2}h^{(2)\mu\nu}\Psi^{(2)}_{\mu\nu} , \label{2.18}
\end{eqnarray}
where the current $\Psi^{(2)}_{\mu\nu}$ is the same energy-momentum tensor (\ref{L1(2)}) and
\begin{eqnarray}
\Psi^{(4)}_{\mu\nu\alpha\beta}=\partial_{(\alpha}F_{\mu}^{\ \rho}\partial_{\beta}F_{\nu)\rho}
-\frac{1}{2}g_{(\mu\nu}\partial^{\lambda}F_{\alpha\sigma}\partial^{\sigma}F_{\beta)\lambda}
-\frac{1}{2}g_{(\mu\nu}\partial_{\alpha}F^{\sigma\rho}\partial_{\beta)}F_{\sigma\rho} .\label{2.19}
\end{eqnarray}
The whole action
\begin{equation}\label{l01}
    \mathcal{L}_{0}(A_{\mu})+\mathcal{L}_{1}(A_{\mu},h^{(2)\mu\nu},h^{(4)\mu\nu\alpha\beta}) ,
\end{equation}
  is invariant with respect to the spin one gauge transformations and the following higher spin transformations
\begin{eqnarray}
&&\delta^{1}A_{\mu}=\epsilon^{\rho\lambda\sigma}\partial_{\rho}\partial_{\lambda}F_{\mu\sigma}
+\frac{1}{2}\partial_{\rho}\epsilon_{\mu\lambda\sigma}\partial^{\lambda}F^{\sigma\rho},\label{2.21}\\
&&\delta^{0}h^{(4)\mu\nu\alpha\beta}=4\partial^{(\mu}\epsilon^{\nu\alpha\beta)},\
\delta_{\epsilon}^{0}h_{\mu}^{\ \mu\alpha\beta}=2\epsilon_{(1)}^{\alpha\beta},\label{2.22}\\
&&\delta^{0}h^{(2)\mu\nu}=2\partial^{(\mu}\epsilon^{\nu)}_{(2)},\ \delta^{0}h_{\mu}^{(2)\mu}=2\epsilon_{(3)} .\label{2.23}
\end{eqnarray}

Therefore we have to prove that like the previously investigated scalar--higher spin coupling case \cite{MR}, the interaction with the spin four gauge field leads to the additional interaction with the lower even spin two field.
To do that according to the previous lesson we start from a spin one gauge invariant ansatz for the spin four transformation of $A_{\mu}$ field
\begin{eqnarray}
\delta_{\epsilon}^{1}A_{\mu}=\epsilon^{\rho\lambda\sigma}\partial_{\rho}\partial_{\lambda}F_{\mu\sigma} .\label{delta4A}
\end{eqnarray}
Thus we have now the following variation  of $\mathcal{L}_{0}$
\begin{eqnarray}
\delta_{\epsilon}^{1}\mathcal{L}_{0}=\delta_{\epsilon}^{1}(-\frac{1}{4}F_{\mu\nu}F^{\mu\nu})
=(\delta_{\epsilon}^{1}A_{\nu})\partial_{\mu}F^{\mu\nu}
=-\partial_{\mu}(\epsilon^{\rho\lambda\sigma}\partial_{\rho}\partial_{\lambda}F_{\nu\sigma})F^{\mu\nu} .
\end{eqnarray}

After some algebra, again neglecting total derivatives and using the Bianchi identity for $F_{\mu\nu}$
\begin{eqnarray}
\partial_{\mu}F_{\nu\lambda}+\partial_{\nu}F_{\lambda\mu}+\partial_{\lambda}F_{\mu\nu}=0,\label{BianciF}
\end{eqnarray}
and taking into account the important relation
\begin{eqnarray}
-\partial^{\mu}\epsilon^{\rho\lambda\sigma}\partial_{\rho}F_{\mu}^{\ \nu}\partial_{\lambda}F_{\sigma\nu}=
    -\partial^{(\mu}\epsilon^{\rho\lambda\sigma)}\partial_{(\rho}F_{\mu}^{\ \nu}\partial_{\lambda}F_{\sigma)\nu}
    +\frac{1}{4}\epsilon_{(1)}^{\lambda\sigma}\partial^{\nu}F_{\mu\lambda}\partial^{\mu}F_{\nu\sigma}\nonumber\\
    -\frac{1}{2}\partial^{\nu}\epsilon^{\rho\lambda\sigma}\partial_{\lambda}F_{\sigma\nu}\partial^{\mu}F_{\mu\rho}
    -\frac{1}{4}\epsilon_{(1)}^{\lambda\sigma}\partial^{\mu}F_{\mu\rho}\partial^{\nu}F_{\nu\sigma},\label{sym4}
\end{eqnarray}
we arrive at the following form of the variation convenient for our analysis
\begin{eqnarray}
\delta_{\epsilon}^{1}\mathcal{L}_{0}&=&-\partial^{(\mu}\epsilon^{\rho\lambda\sigma)}\partial_{(\rho}F_{\mu}^{\ \nu}\partial_{\lambda}F_{\sigma)\nu}+\frac{1}{4}\epsilon_{(1)}^{\lambda\sigma}\partial^{\nu}F_{\mu\lambda}\partial^{\mu}F_{\nu\sigma}
+\frac{1}{4}\epsilon_{(1)}^{\lambda\sigma}\partial_{\lambda}F_{\mu\nu}\partial_{\sigma}F^{\mu\nu}\nonumber\\
&-&\partial_{\lambda}(\epsilon_{(1)}^{\lambda\sigma}F_{\mu\sigma})\partial_{\nu}F^{\nu\mu}
-\frac{1}{4}\epsilon_{(1)}^{\lambda\sigma}\partial^{\mu}F_{\mu\lambda}\partial^{\nu}F_{\nu\sigma}
-\frac{1}{2}\partial^{\rho}\epsilon^{\nu\lambda\sigma}\partial_{\lambda}F_{\sigma\rho}\partial^{\mu}F_{\mu\nu}\nonumber\\
&+&\partial^{(\mu}\epsilon_{(2)}^{\nu)}F_{\mu\sigma}F_{\nu}^{\ \sigma}
-\frac{1}{4}\epsilon_{(3)}F_{\mu\nu}F^{\mu\nu}.\label{delta4L0sym}
\end{eqnarray}
Returning to the gauge variation of the spin four field (\ref{h4}) we notice that all terms in the first line of (\ref{delta4L0sym}) and the first two terms in the second line can be integrated to the interaction terms.
The last term in the second line is proportional to the free field equations but is not integrable, so we can cancel this term only by changing the initial variation of $A_{\mu}$ (\ref{delta4A}). The modified form of (\ref{delta4A}) is
\begin{eqnarray}
\delta_{\epsilon}^{1}A_{\mu}=\epsilon^{\rho\lambda\sigma}\partial_{\rho}\partial_{\lambda}F_{\mu\sigma}
+\frac{1}{2}\partial_{\rho}\epsilon_{\mu\lambda\sigma}\partial^{\lambda}F^{\sigma\rho}.
\end{eqnarray}
Therefore
\begin{eqnarray}
\mathcal{L}_{1}&=&\frac{1}{4}h^{(4)\mu\rho\lambda\sigma}\partial_{(\rho}F_{\mu}^{\ \nu}\partial_{\lambda}F_{\sigma)\nu}-\frac{1}{8}h_{\rho}^{(4)\rho\lambda\sigma}\partial^{\nu}F_{\mu\lambda}\partial^{\mu}F_{\nu\sigma}
-\frac{1}{8}h_{\rho}^{(4)\rho\lambda\sigma}\partial_{\lambda}F_{\mu\nu}\partial_{\sigma}F^{\mu\nu}\nonumber\\
&+&\partial_{\lambda}(\frac{1}{2}h_{\rho}^{(4)\rho\lambda\sigma}F_{\mu\sigma})\partial_{\nu}F^{\nu\mu}
+\frac{1}{8}h_{\rho}^{(4)\rho\lambda\sigma}\partial^{\mu}F_{\mu\lambda}\partial^{\nu}F_{\nu\sigma}\nonumber\\
&-&\frac{1}{2}h^{(2)\mu\nu}F_{\mu\sigma}F_{\nu}^{\ \sigma}
+\frac{1}{8}h^{(2)\rho}_{\rho}F_{\mu\nu}F^{\mu\nu}.\label{delta4L0sym1}
\end{eqnarray}

But the two terms in the second line are proportional to the equation of motion for the initial Lagrangian (\ref{L0}), hence they are not physical and can be removed by the following field redefinition
\begin{eqnarray}
A_{\mu} \rightarrow A_{\mu}-\partial_{\lambda}(\frac{1}{2}h_{\alpha}^{\ \alpha\lambda\sigma}F_{\mu\sigma})
-\frac{1}{8}h_{\ \alpha\mu\sigma}^{\alpha}\partial_{\beta}F^{\beta\sigma} .
\end{eqnarray}
So we can drop the second line of (\ref{delta4L0sym1}).

Another novelty in (\ref{delta4L0sym1}) in comparison with the previous case is the third line of (\ref{delta4L0sym}). Comparing with (\ref{211}) we see that we can integrate these two terms introducing an additional spin two field coupling and compensate the first and third line introducing the linearized Lagrangian (\ref{2.18})  for the coupling of the electromagnetic field to the spin four and spin two fields with the set of higher spin field transformations (\ref{2.21})-(\ref{2.23}).

Therefore we proved that the interaction with the spin four gauge field leads to the additional interaction with the lower even spin two field.

\section{Generalization to the 2-2-4 and 2-2-6 interactions}
\setcounter{equation}{0}
In this section we turn to the spin two field as a lower spin field in the construction of the higher spin gauge invariant interactions with spin 4 and spin 6 gauge potentials. And again we want to keep manifest the lower spin two gauge invariance.

So proceeding similarly as in the previous section we start from the free spin two Pauli-Fierz Lagrangian \cite{PF}
\begin{equation}\label{3.1}
\mathcal{L}_{0}(h^{(2)}_{\mu\nu})=\frac{1}{2}\partial_{\mu}h^{(2)}_{\alpha\beta}\partial^{\mu}h^{(2)\alpha\beta}
-\partial_{\alpha}h^{(2)\alpha\beta}\partial_{\mu}h^{(2)\mu}_{\beta}+\partial_{\mu}h^{(2)\alpha}_{\alpha}\partial_{\beta}h^{(2)\beta\mu}
-\frac{1}{2}\partial_{\mu}h^{(2)\alpha}_{\alpha}\partial^{\mu}h^{(2)\beta}_{\beta} ,
\end{equation}
and try to solve the following Noether's equations, either
\begin{equation}
    \delta_{\varepsilon}^{1}\mathcal{L}_{0}(h^{(2)}_{\mu\nu})
    +\delta_{\varepsilon}^{0}\mathcal{L}_{1}(h^{(2)}_{\mu\nu},h^{(4)\alpha\beta\lambda\rho} )=0 ,\label{3.2} \end{equation}
or
\begin{equation}
    \delta_{\varepsilon}^{1}\mathcal{L}_{0}(h^{(2)}_{\mu\nu})
    +\delta_{\varepsilon}^{0}\mathcal{L}_{1}(h^{(2)}_{\mu\nu},h^{(4)\alpha\beta\lambda\rho},h^{(6)\mu\nu\alpha\beta\lambda\rho} )=0 .\label{3.2(1)}
\end{equation}
Again  we present first the final result for the 2-2-4 gauge invariant interaction
\begin{eqnarray}
&&\mathcal{L}_{1}(h^{(2)}_{\mu\nu},h^{(4)}_{\alpha\beta\mu\nu})=\frac{1}{4}h^{(4)\alpha\beta\mu\nu}
\Psi^{(4)}_{(\Gamma)\alpha\beta\mu\nu}(h^{(2)}_{\mu\nu})\nonumber\\
&&=\frac{1}{4}h^{(4)\alpha\beta\mu\nu}\Gamma_{\alpha\beta,\rho\sigma}\Gamma_{\mu\nu,}^{\ \ \ \rho\sigma}
-\frac{1}{6}h^{(4)\alpha\mu\nu}_{\alpha}\Gamma_{\mu}^{\ \rho,\sigma\lambda}\Gamma_{\nu\rho,\sigma\lambda} , \label{3.3}
\end{eqnarray}
with the following gauge transformations
\begin{eqnarray}
&&\delta_{\epsilon}h^{(2)}_{\mu\nu}=\epsilon^{\rho\lambda\sigma}\partial_{\rho}\Gamma_{\lambda\sigma,\mu\nu}
-\partial_{\rho}\epsilon_{\lambda\sigma(\mu}\Gamma_{\nu)}^{\ \rho,\lambda\sigma} ,\label{3.5(1)}\\
&&\delta_{\epsilon}^{0}h^{(4)\mu\rho\lambda\sigma}=4\partial^{(\mu}\epsilon^{\rho\lambda\sigma)},\quad
\delta_{\epsilon}^{0}h_{\rho}^{(4) \rho\lambda\sigma}=2\epsilon_{(1)}^{\lambda\sigma}.\label{3.5(2)}
\end{eqnarray}
The final result for the 2-2-6 case correspondingly looks like
\begin{eqnarray}
&& \mathcal{L}_{1}(h^{(2)}, h^{(4)}, h^{(6)})
=-\frac{1}{6}h^{(6)\alpha\beta\mu\nu\lambda\rho}\Psi^{(6)}_{(\Gamma)\alpha\beta\mu\nu\lambda\rho}
+\frac{1}{4}h^{(4)\alpha\beta\mu\nu}\Psi^{(4)}_{(\Gamma)\alpha\beta\mu\nu}\nonumber\\
&& =-\frac{1}{6}h^{(6)\alpha\beta\mu\nu\lambda\rho}\partial_{\alpha}\Gamma_{\beta\mu,}^{\quad\,\,\,\,\,\sigma\delta}
\partial_{\nu}\Gamma_{\lambda\rho,\sigma\delta}
+\frac{1}{6}h_{\alpha}^{(6)\alpha\mu\nu\lambda\rho}\partial_{\mu}\Gamma_{\nu}^{\ \kappa,\sigma\delta}\partial_{\lambda}\Gamma_{\rho\kappa,\sigma\delta}
\nonumber\\&&+\frac{1}{12}h_{\alpha}^{(6)\alpha\mu\nu\lambda\rho}\partial^{\kappa}\Gamma_{\mu\nu,}^{\quad\sigma\delta}
\partial_{\sigma}\Gamma_{\lambda\rho),\kappa\delta}
+\frac{1}{4}h^{(4)\alpha\beta\mu\nu}\Gamma_{\alpha\beta,\rho\sigma}\Gamma_{\mu\nu,}^{\ \ \ \rho\sigma}
-\frac{1}{6}h^{(4)\alpha\mu\nu}_{\alpha}\Gamma_{\mu}^{\ \rho,\sigma\lambda}\Gamma_{\nu\rho,\sigma\lambda} .\quad\quad\label{3.7(1)}
\end{eqnarray}
This formula together with the corrected gauge transformation
\begin{eqnarray}
&&\delta^{1}_{\epsilon}h^{(2)}_{\alpha\beta}=\epsilon^{\mu\nu\rho\lambda\sigma}
\partial_{\mu}\partial_{\nu}\partial_{\rho}\Gamma_{\lambda\sigma,\alpha\beta}
-\frac{4}{3}\partial^{\rho}\epsilon^{\,\,\,\,\mu\nu\lambda\sigma}_{\alpha}
\partial_{\lambda}\partial_{\sigma}\Gamma_{\beta\rho ,\mu\nu}+\frac{1}{3}\partial^{\rho}\partial^{\lambda}
  \epsilon^{\,\,\,\,\,\,\,\mu\nu\sigma}_{\alpha\beta}\partial_{\sigma}\Gamma_{\rho\lambda ,\mu\nu} ,\quad\quad \label{3.8(1)}\\
  &&\delta^{0}_{\epsilon}h^{(6)\mu\nu\alpha\beta\sigma\rho}=6 \partial^{(\mu}\epsilon^{\nu\alpha\beta\sigma\rho)}(x) ,\quad \delta^{0}_{\epsilon}h^{(6)\mu\alpha\beta\sigma\rho}_{\mu}=2\epsilon^{\alpha\beta\sigma\rho}_{(1)} .\label{3.9(1)}\\
&&\delta_{\epsilon}^{0}h^{(4)\mu\rho\lambda\sigma}=4\partial^{(\mu}\epsilon^{\rho\lambda\sigma)}_{(2)},\quad
\delta_{\epsilon}^{0}h_{\rho}^{(4) \rho\lambda\sigma}=2\epsilon_{(3)}^{\lambda\sigma} \label{3.10(1)}
\end{eqnarray}
solves completely Noether's equation (\ref{3.2(1)}).

$\Gamma_{\lambda\sigma,\mu\nu}$ here is the spin two gauge invariant symmetrized linearized Riemann curvature
\begin{eqnarray}
&&\Gamma_{\alpha\beta,\mu\nu}=\frac{1}{2}(R_{\alpha\mu,\beta\nu}+R_{\beta\mu,\alpha\nu}) ,\\
&& \Gamma_{(\alpha\beta,\mu)\nu}=0 ,\label{3.5}
\end{eqnarray}
introduced by de Witt and Freedman for higher spin gauge fields together with the higher spin generalization of the
Christoffel symbols \cite{DF}. This symmetrized curvature is more convenient for the construction of an interaction with symmetric tensors.
The corresponding Ricci tensor (Fronsdal operator for higher spin generalization) and scalar can be defined in the usual manner using traces
\begin{eqnarray}
&&\mathcal{F}_{\mu\nu}=\Gamma_{\mu\nu,\lambda}^{\lambda}=\Box h^{(2)}_{\mu\nu}-2\partial_{(\mu}\partial^{\alpha}h^{(2)}_{\nu)\alpha}+\partial_{\mu}\partial_{\nu}h^{(2)\alpha}_{\alpha} , \\
&&\mathcal{F}=\mathcal{F}_{\mu}^{\mu}=2(\Box h^{(2)\mu}_{\mu}-\partial_{\mu}\partial_{\nu}h^{(2)\mu\nu}).
\end{eqnarray}
In terms of these objects the Bianchi identities can be written as
\begin{eqnarray}
  &&\partial_{\lambda}\Gamma_{\mu\nu,\alpha\beta}=\partial_{(\mu}\Gamma_{\nu)\lambda,\alpha\beta}
  +\partial_{(\alpha}\Gamma_{\beta)\lambda,\mu\nu} , \label{3.8}\\
  &&\partial_{\lambda}\mathcal{F}_{\alpha\beta}=\partial^{\mu}\Gamma_{\mu\lambda,\alpha\beta}
  +\partial_{(\alpha}\mathcal{F}_{\beta)\lambda} , \label{3.9}\\
  && \partial^{\lambda}\mathcal{F}_{\lambda\mu}=\frac{1}{2}\partial_{\mu}\mathcal{F}^{\alpha}_{\alpha}.\label{3.10}
\end{eqnarray}

So to prove (\ref{3.3})-(\ref{3.5(2)}) we introduce the following starting ansatz for the spin four transformation of the spin two field
\begin{eqnarray}
\delta^{1}_{\epsilon}h^{(2)}_{\mu\nu}=\epsilon^{\rho\lambda\sigma}\partial_{\rho}\Gamma_{\lambda\sigma,\mu\nu} ,\label{3.11}
\end{eqnarray}
Then a variation of (\ref{3.1}) with respect to (\ref{3.3}) is
\begin{eqnarray}
  && \delta^{1}_{\epsilon}\mathcal{L}_{0}(h^{(2)}_{\mu\nu})=\frac{\delta \mathcal{L}_{0}}{\delta h^{(2)}_{\mu\nu}}\delta^{1}_{\epsilon}h^{(2)}_{\mu\nu}=-(\mathcal{F}^{\mu\nu}-\frac{1}{2}g^{\mu\nu}\mathcal{F})
  \epsilon^{\rho\lambda\sigma}\partial_{\rho}\Gamma_{\lambda\sigma,\mu\nu} .
\end{eqnarray}
To integrate it and solve the equation (\ref{3.2}) we submit to the following strategy:

1) First we perform a partial integration and use the Bianchi identity (\ref{3.9}) to lift the variation to a curvature square term.

2) Then we make a partial integration again and rearrange indices using (\ref{3.5}) and (\ref{3.8}) to extract an integrable part.

3) Symmetrizing expressions in this way we classify terms as
\begin{itemize}
  \item integrable
  \item integrable and subjected to field redefinition (proportional to the free field equation of motion)
  \item non integrable but reducible by deformation of the initial ansatz for the gauge transformation (again proportional to the free field equation of motion)
\end{itemize}

Then if no other terms remain we can construct our interaction together with the corrected first order transformation.
Following this strategy after some fight with formulas we win the battle obtaining the following expression
\begin{eqnarray}
\delta^{1}_{\epsilon}\mathcal{L}_{0}(h^{(2)}_{\mu\nu})&=&-\partial^{(\alpha}\epsilon^{\beta\mu\nu)}
(\Psi^{(4)}_{(\Gamma)\alpha\beta\mu\nu}-\Psi^{(4)}_{(\mathcal{F})\alpha\beta\mu\nu})
\nonumber\\&-&\epsilon^{\mu\nu}_{(1)}\Gamma_{\mu\nu,\alpha\beta}\frac{\delta \mathcal{L}_{0}}{\delta  h^{(2)}_{\alpha\beta}} +\partial^{\rho}\epsilon^{\,\,\,\mu\nu}_{\alpha}\Gamma_{\beta\rho ,\mu\nu}\frac{\delta \mathcal{L}_{0}}{\delta  h^{(2)}_{\alpha\beta}} , \label{3.12}
\end{eqnarray}
where
\begin{eqnarray}
&&\Psi^{(4)}_{(\Gamma)\alpha\beta\mu\nu}=\Gamma_{(\alpha\beta,}^{\,\,\,\,\,\,\,\,\,\,\,\,\rho\sigma}\Gamma_{\mu\nu),\rho\sigma}
-\frac{2}{3}g_{(\alpha\beta}\Gamma_{\mu}^{\ \rho,\sigma\lambda}\Gamma_{\nu)\rho,\sigma\lambda} ,\label{3.13}\\
&&\Psi^{(4)}_{(\mathcal{F})\alpha\beta\mu\nu}=\mathcal{F}_{(\alpha\beta}\mathcal{F}_{\mu\nu)}
-g_{(\alpha\beta}\mathcal{F}_{\mu}^{\sigma}\mathcal{F}_{\nu)\sigma}=-\frac{\delta \mathcal{L}_{0}}{\delta  h^{(2)(\alpha\beta}}\mathcal{F}_{\mu\nu)}
+g_{(\alpha\beta}\frac{\delta \mathcal{L}_{0}}{\delta  h^{(2)\mu}_{\sigma}}\mathcal{F}_{\nu)\sigma} ,\quad\quad\label{3.14}\\
&&\frac{\delta \mathcal{L}_{0}}{\delta  h^{(2)\alpha\beta}}=-\mathcal{F}_{\alpha\beta}+\frac{1}{2}g_{\alpha\beta}\mathcal{F} .\label{3.15}
\end{eqnarray}
So we see immediately that in (\ref{3.12}) only the last term of the second line is not integrable but proportional to the equation of motion and can be dropped by the correction (\ref{3.5(1)}) to the initial gauge transformation (\ref{3.11}). Other terms of (\ref{3.12}) can be integrated to
\begin{eqnarray}
&&\mathcal{L}_{1}(h^{(2)}_{\mu\nu},h^{(4)}_{\alpha\beta\mu\nu})=\frac{1}{4}h^{(4)\alpha\beta\mu\nu}
\left(\Psi^{(4)}_{(\Gamma)\alpha\beta\mu\nu}(h^{(2)}_{\mu\nu})-\Psi^{(4)}_{(\mathcal{F})\alpha\beta\mu\nu}\right)
+\frac{1}{2}h^{(4)\alpha\mu\nu}_{\alpha}\Gamma_{\mu\nu,\alpha\beta}\frac{\delta \mathcal{L}_{0}}{\delta  h^{(2)}_{\alpha\beta}}\label{3.24(1)} .\nonumber\\
\end{eqnarray}

On the other hand taking into account (\ref{3.14}) and (\ref{3.15}) we can compensate $\Psi^{(4)}_{(\mathcal{F})}$ and the last term in (\ref{3.24(1)}) by the following field redefinition
\begin{eqnarray}
h^{(2)}_{\mu\nu} \rightarrow h^{(2)}_{\mu\nu} -\frac{1}{2}h^{(4)\alpha\lambda\sigma}_{\alpha}\Gamma_{\lambda\sigma,\mu\nu}
-\frac{1}{4}h^{(4)\alpha\lambda}_{\mu\nu}\mathcal{F}_{\alpha\lambda}
+\frac{1}{4}h^{(4)\alpha\lambda}_{\ \ \ \alpha(\mu}\mathcal{F}_{\nu)\lambda} .
\end{eqnarray}
Thus after field redefinition we arrive at the 2-2-4 gauge invariant interaction (\ref{3.3}) with the gauge transformations (\ref{3.5(1)}), (\ref{3.5(2)}).

Now in possession of knowledge about the 2-2-4 interaction we start to construct the most nontrivial interaction in this article between spin 2 and spin 6 gauge fields (\ref{3.7(1)})-(\ref{3.10(1)}). We would like to check the appearance of the 2-2-4 coupling during the construction of 2-2-6 which we expect from the analogy with the scalar case considered in \cite{MMR, MR} and the 1-1-4 case considered in the previous section.

To proceed we have to solve the following initial Noether's equation
\begin{equation}\label{3.20}
    \delta_{\varepsilon}^{1}\mathcal{L}_{0}(h^{(2)}_{\mu\nu})
    +\delta_{\varepsilon}^{0}\mathcal{L}_{1}(h^{(2)}_{\mu\nu},h^{(6)}_{\alpha\beta\lambda\rho\sigma\delta} )=0 ,
\end{equation}
with a starting ansatz for the spin 6 first order gauge transformation for the spin 2 field:
\begin{eqnarray}
\delta^{1}_{\epsilon}h^{(2)}_{\mu\nu}(x)=\epsilon^{\alpha\beta\rho\lambda\sigma}(x)
\partial_{\alpha}\partial_{\beta}\partial_{\rho}\Gamma_{\lambda\sigma,\mu\nu}(x) ,\label{3.21}
\end{eqnarray}
and the standard zero order gauge transformation for the spin 6 gauge field
\begin{eqnarray}
 &&\delta^{0}_{\epsilon}h^{(6)\mu\nu\alpha\beta\sigma\rho}=6 \partial^{(\mu}\epsilon^{\nu\alpha\beta\sigma\rho)}(x) ,\label{3.22}\\
 && \delta^{0}_{\epsilon}h^{(6)\mu\alpha\beta\sigma\rho}_{\mu}=2\epsilon^{\alpha\beta\sigma\rho}_{(1)} .\label{3.23}
\end{eqnarray}
First of all we have to transform the variation
\begin{eqnarray}
  && \delta_{\varepsilon}^{1}\mathcal{L}_{0}(h^{(2)}_{\mu\nu})=-(\mathcal{F}^{\mu\nu}-\frac{1}{2}g^{\mu\nu}\mathcal{F})
  \epsilon^{\alpha\beta\rho\lambda\sigma}
\partial_{\alpha}\partial_{\beta}\partial_{\rho}\Gamma_{\lambda\sigma,\mu\nu} ,\label{3.24}
\end{eqnarray}
into a form convenient for integration. Following the same strategy as before in the 2-2-4 case, using many times partial integration and Bianchi identities (\ref{3.5}), (\ref{3.8})-(\ref{3.10}), we obtain after tedious but straightforward calculations
\begin{eqnarray}
  &&\delta_{\varepsilon}^{1}\mathcal{L}_{0}(h^{(2)}_{\mu\nu})=\partial^{(\alpha}\epsilon^{\beta\mu\nu\lambda\rho)}
\Psi^{(6)}_{(\Gamma)\alpha\beta\mu\nu\lambda\rho}-\partial^{(\alpha}\epsilon^{\beta\mu\nu)}_{(2)}\Psi^{(4)}_{(\Gamma)\alpha\beta\mu\nu} \nonumber\\
  &&+\frac{4}{3}\partial^{\rho}\epsilon^{\,\,\,\,\mu\nu\lambda\sigma}_{\alpha}\partial_{\lambda}\partial_{\sigma}\Gamma_{\beta\rho ,\mu\nu}\frac{\delta \mathcal{L}_{0}}{\delta  h^{(2)}_{\alpha\beta}}-\frac{1}{3}\partial^{\rho}\partial^{\lambda}
  \epsilon^{\,\,\,\,\,\,\,\mu\nu\sigma}_{\alpha\beta}\partial_{\sigma}\Gamma_{\rho\lambda ,\mu\nu}\frac{\delta \mathcal{L}_{0}}{\delta  h^{(2)}_{\alpha\beta}}\nonumber\\
  &&- R_{int}^{\mu\nu}(\Gamma,\mathcal{F},\epsilon)\frac{\delta\mathcal{L}_{0}}{\delta h^{(2)}_{\mu\nu}} ,\label{3.25}
\end{eqnarray}
where
\begin{eqnarray}
 &&  \Psi^{(6)}_{(\Gamma)\alpha\beta\mu\nu\lambda\rho}=
  \partial_{(\alpha}\Gamma_{\beta\mu,}^{\quad\,\,\,\,\,\sigma\delta}\partial_{\nu}\Gamma_{\lambda\rho),\sigma\delta}
-g_{(\alpha\beta}\partial_{\mu}\Gamma_{\nu}^{\ \kappa,\sigma\delta}\partial_{\lambda}\Gamma_{\rho)\kappa,\sigma\delta}\nonumber\\
&&\qquad\qquad \quad -\frac{1}{2}g_{(\alpha\beta}\partial^{\kappa}\Gamma_{\mu\nu ,}^{\quad\sigma\delta}\partial_{\sigma}\Gamma_{\lambda\rho),\kappa\delta} ,\quad\quad\quad\quad\\
   &&\Psi^{(4)}_{(\Gamma)\alpha\beta\mu\nu}=\Gamma_{(\alpha\beta,}^{\quad\quad\rho\sigma}\Gamma_{\mu\nu),\rho\sigma}
-\frac{2}{3}g_{(\alpha\beta}\Gamma_{\mu}^{\ \rho,\sigma\lambda}\Gamma_{\nu)\rho,\sigma\lambda} ,
\end{eqnarray}
and $R_{int}^{\mu\nu}(\Gamma,\mathcal{F},\epsilon)\frac{\delta\mathcal{L}_{0}}{\delta h^{(2)}_{\mu\nu}}$ are remaining integrable terms proportional to the equation of motion. Indeed the symmetric tensor $R_{int}^{\mu\nu}(\Gamma,\mathcal{F})$ is expressed through the only integrable combinations of derivatives of the gauge parameter
\begin{eqnarray}
  R_{int}^{\mu\nu}(\Gamma,\mathcal{F},\epsilon)&=&
  \epsilon^{\alpha\beta\lambda\delta}_{(1)}\partial_{\alpha}\partial_{\beta}
  \Gamma_{\lambda\delta,}^{\quad \mu\nu}-\frac{1}{3}\partial^{\lambda}\epsilon^{\alpha\beta\delta(\mu}_{(1)}\partial_{\alpha}\Gamma_{\,\,\lambda,\beta\delta}^{\nu)}
  +\partial_{\lambda}\left[\partial^{(\lambda}\epsilon^{\alpha\beta\delta\mu\nu)}\partial_{\alpha}\mathcal{F}_{\beta\delta}\right] \nonumber\\
  &-&\frac{2}{3}\partial_{\lambda}\left[\epsilon^{\lambda\alpha\mu\nu}_{(1)}
  \partial_{\alpha}\mathcal{F}\right]+\frac{1}{6}\epsilon^{\alpha\beta\mu\nu}_{(1)}\partial_{\alpha}\partial_{\beta}\mathcal{F}
  +\partial^{(\alpha}\epsilon^{\beta\mu\nu)}_{(2)}\mathcal{F}_{\alpha\beta} + \frac{5}{3}\partial^{\alpha}\epsilon^{\beta\lambda\mu\nu}_{(1)}\partial_{\lambda}\mathcal{F}_{\alpha\beta} \nonumber\\
  &-&\frac{5}{3}\partial_{\lambda}\left[\epsilon^{\lambda\alpha\beta(\mu}_{(1)}\partial_{\alpha}\mathcal{F}_{\beta}^{\nu}\right] + \frac{1}{6}\Box\epsilon^{\alpha\beta\mu\nu}_{(1)}\mathcal{F}_{\alpha\beta}-\frac{1}{6}\partial^{\lambda}
  \epsilon^{\alpha\beta\mu\nu}_{(1)}\partial_{\lambda}\mathcal{F}_{\alpha\beta}-\frac{1}{2} \epsilon^{\alpha(\mu}_{(3)}\mathcal{F}^{\nu)}_{\alpha} .\nonumber \label{3.29}\\
\end{eqnarray}
The second line in (\ref{3.25}) is not integrable and therefore can be cancelled by the following deformation of the initial ansatz for the transformation (\ref{3.21})
\begin{eqnarray}
\delta^{1}_{\epsilon}h^{(2)}_{\alpha\beta}=\epsilon^{\mu\nu\rho\lambda\sigma}
\partial_{\mu}\partial_{\nu}\partial_{\rho}\Gamma_{\lambda\sigma,\alpha\beta}-\frac{4}{3}\partial^{\rho}\epsilon^{\,\,\,\,\mu\nu\lambda\sigma}_{\alpha}
\partial_{\lambda}\partial_{\sigma}\Gamma_{\beta\rho ,\mu\nu}+\frac{1}{3}\partial^{\rho}\partial^{\lambda}
  \epsilon^{\,\,\,\,\,\,\,\mu\nu\sigma}_{\alpha\beta}\partial_{\sigma}\Gamma_{\rho\lambda ,\mu\nu} .\label{3.30}
\end{eqnarray}

Then substituting into (\ref{3.29})  $\partial^{(\lambda}\epsilon^{\alpha\beta\delta\mu\nu)}$ with $\frac{1}{6}h^{(6)\lambda\alpha\beta\delta\mu\nu}$ , $\partial^{(\alpha}\epsilon^{\beta\mu\nu)}_{(2)}$ with $\frac{1}{4}h^{(4)\alpha\beta\mu\nu}$, and correspondingly $2\epsilon^{\alpha\beta\mu\nu}_{(1)}$ and $2\epsilon^{\alpha\beta}_{(3)}$ with their traces, we can integrate  the first and third  line of (\ref{3.25}) to
\begin{eqnarray}
&& \mathcal{L}_{1}(h^{(2)}, h^{(4)}, h^{(6)})
=-\frac{1}{6}h^{(6)\alpha\beta\mu\nu\lambda\rho}\Psi^{(6)}_{(\Gamma)\alpha\beta\mu\nu\lambda\rho}
+\frac{1}{4}h^{(4)\alpha\beta\mu\nu}\Psi^{(4)}_{(\Gamma)\alpha\beta\mu\nu}\nonumber\\
&& +R_{int}^{\mu\nu}(\Gamma,\mathcal{F},h^{(6)},h^{(4)})\frac{\delta\mathcal{L}_{0}}{\delta h^{(2)}_{\mu\nu}} \label{3.25(1)}
\end{eqnarray}
where
\begin{eqnarray}
  &&R_{int}^{\mu\nu}(\Gamma,\mathcal{F},h^{(6)},h^{(4)})=
  \frac{1}{2}h_{\rho}^{(6)\rho\alpha\beta\lambda\delta}\partial_{\alpha}\partial_{\beta}
  \Gamma_{\lambda\delta,}^{\quad \mu\nu}-\frac{1}{6}\partial^{\lambda}h_{\rho}^{(6)\rho\alpha\beta\delta(\mu}\partial_{\alpha}\Gamma_{\,\,\lambda,\beta\delta}^{\nu)}
  +\partial_{\lambda}\left[\frac{1}{6}h^{(6)\lambda\alpha\beta\delta\mu\nu}\partial_{\alpha}\mathcal{F}_{\beta\delta}\right] \nonumber\\
  &&-\frac{2}{6}\partial_{\lambda}\left[h_{\rho}^{(6)\rho\lambda\alpha\mu\nu}
  \partial_{\alpha}\mathcal{F}\right]+\frac{1}{12}h_{\rho}^{(6)\rho\alpha\beta\mu\nu}\partial_{\alpha}\partial_{\beta}\mathcal{F}
  +\frac{1}{4}h^{(4)\alpha\beta\mu\nu}\mathcal{F}_{\alpha\beta} + \frac{5}{6}\partial^{\alpha}h_{\rho}^{(6)\rho\beta\lambda\mu\nu}\partial_{\lambda}\mathcal{F}_{\alpha\beta} \nonumber\\
  &&-\frac{5}{6}\partial_{\lambda}\left[h_{\rho}^{(6)\rho\lambda\alpha\beta(\mu}\partial_{\alpha}\mathcal{F}_{\beta}^{\nu)}\right] + \frac{1}{12}\Box h_{\rho}^{(6)\rho\alpha\beta\mu\nu}\mathcal{F}_{\alpha\beta}-\frac{1}{12}\partial^{\lambda}
  h_{\rho}^{(6)\rho\alpha\beta\mu\nu}\partial_{\lambda}\mathcal{F}_{\alpha\beta}-\frac{1}{4} h_{\rho}^{(4)\rho\alpha(\mu}\mathcal{F}^{\nu)}_{\alpha} .\nonumber \label{3.29(1)}\\
\end{eqnarray}
Now we define a field redefinition for $h^{(2)\mu\nu}$
\begin{equation}
    h^{(2)\mu\nu} \rightarrow h^{(2)\mu\nu} - R_{int}^{\mu\nu}(\Gamma,\mathcal{F},h^{(6)},h^{(4)}) ,
\end{equation}
using which we can drop the last term in (\ref{3.25(1)}).

Thus we arrive at the promised result that the 2-2-6 interaction automatically includes also the 2-2-4 interaction constructed above, and the corresponding trilinear interaction Lagrangian is (\ref{3.7(1)}). This formula together with the corrected gauge transformations (\ref{3.8(1)})-(\ref{3.10(1)}) solves completely Noether's equation (\ref{3.2(1)}).

Finally note that these interactions should reproduce the flat space limit of the Fradkin-Vasiliev type nonlinear interactions \cite{Vasiliev} constructed in an $AdS$ background. For some other vertices  i.e. 2-s-s and 1-s-s with additional nonabelian symmetry such construction and connection with Fradkin-Vasiliev formalism can be found in \cite{boulanger}, where authors used BRST-cohomological approach.

\section{2s-s-s interaction Lagrangian}\setcounter{equation}{0}
\quad The most elegant and convenient  way of handling symmetric tensors such as $h^{(s)}_{\mu_1\mu_2...\mu_s}(z)$ is by
contracting it with the $s$'th tensorial power of a vector $a^{\mu}$ of the tangential space at
the base point $z$ \cite{MR1}-\cite{MR5}
\begin{equation}
h^{(s)}(z;a) = \sum_{\mu_{i}}(\prod_{i=1}^{s} a^{\mu_{i}})h^{(s)}_{\mu_1\mu_2...\mu_s}(z) .
\label{2.5}
\end{equation}
In this way we obtain a homogeneous polynomial in the vector $a^{\mu}$ of degree $s$.
In this formalism the symmetrized gradient, trace and divergence are\footnote{To distinguish easily between "a" and "z" spaces we introduce for space-time derivatives $\frac{\partial}{\partial z^{\mu}}$ the notation $\nabla_{\mu}$ and as before we will admit integration everywhere where it is necessary (we work with a Lagrangian as with an action) and therefore we will neglect all space-time total derivatives when making a partial integration}
\begin{eqnarray}
&&Grad:h^{(s)}(z;a)\Rightarrow Gradh^{(s+1)}(z;a) = (a\nabla)h^{(s)}(z;a) , \\
&&Tr:h^{(s)}(z;a)\Rightarrow Trh^{(s-2)}(z;a) = \frac{1}{s(s-1)}\Box_{a}h^{(s)}(z;a) ,\\
&&Div:h^{(s)}(z;a)\Rightarrow Divh^{(s-1)}(z;a) = \frac{1}{s}(\nabla\partial_{a})h^{(s)}(z;a) .
\end{eqnarray}
The gauge variation of a spin $s$ field is
\begin{eqnarray}\label{4.5}
\delta h^{(s)}(z;a)=s (a\nabla)\epsilon^{(s-1)}(z;a) ,
\end{eqnarray}
with traceless gauge parameter
\begin{eqnarray}
\Box_{a}\epsilon^{(s-1)}(z;a)=0 ,\label{4.6}
\end{eqnarray}
for the double traceless gauge field
\begin{eqnarray}
\Box_{a}^{2}h^{(s)}(z;a)=0 .
\end{eqnarray}

We will use the deWit-Freedman curvature and Cristoffel symbols \cite{DF, MR6}. We contract them with the degree $s$
tensorial power of one tangential vector $a^{\mu}$ in the first set of s indices  and with a similar tensorial power of another tangential vector $b^{\nu}$ in its second set. The deWit-Freedman curvature and n-th Cristoffel symbol  are then written as
\begin{eqnarray}
&&\Gamma^{(s)}(z;b,a):\qquad\Gamma^{(s)}(z; b,\lambda a) = \Gamma^{(s)}(z;\lambda b, a)= \lambda^{s}\Gamma^{(s)}(z;b, a) ,\\
&&\Gamma^{(s)}_{(n)}(z;b,a):\qquad\Gamma^{(s)}_{(n)}(z;b,\lambda a) = \lambda^{s}\Gamma^{(s)}_{(n)}(z;b, a) ,\\
&&\quad\quad\quad\quad\quad\quad\quad\,\,\,\,\Gamma^{(s)}_{(n)}(z; \lambda b, a)=\lambda^{n}\Gamma^{(s)}_{(n)}(z;b, a) ,\\
&&\Gamma^{(s)}(z;b,a)=\Gamma^{(s)}_{(n)}(z;b,a)|_{n=s} .
\label{2.6}
\end{eqnarray}
Next we introduce the notation $*_a, *_b$ for a contraction in the symmetric spaces of indices $a$ or $b$
\begin{eqnarray}
  *_{a}&=&\frac{1}{(s!)^{2}} \prod^{s}_{i=1}\overleftarrow{\partial}^{\mu_{i}}_{a}\overrightarrow{\partial}_{\mu_{i}}^{a} .
   \label{4.12}
\end{eqnarray}
All required manipulations in the framework of this formalism are discussed in the Appendix of this paper. Here we will only present
Fronsdal's Lagrangian in terms of these conventions:
\begin{equation}\label{4.42(1)}
 \mathcal{L}_{0}(h^{(s)}(a))=-\frac{1}{2}h^{(s)}(a)*_{a}\mathcal{F}^{(s)}(a)
    +\frac{1}{8s(s-1)}\Box_{a}h^{(s)}(a)*_{a}\Box_{a}\mathcal{F}^{(s)}(a) .
\end{equation}
where $\mathcal{F}^{(s)}(z;a)$ is so called Fronsdal tensor
\begin{eqnarray}
\mathcal{F}^{(s)}(z;a)=\Box h^{(s)}(z;a)-(a\nabla)(\nabla\partial_{a})h^{(s)}(z;a)
+\frac{1}{2}(a\nabla)^{2}\Box_{a}h^{(s)}(z;a) \quad\label{4.32(1)}
\end{eqnarray}

To obtain the equation of motion we vary (\ref{4.42(1)}) and obtain
\begin{equation}\label{4.45}
     \delta\mathcal{L}_{0}(h^{(s)}(a))=-(\mathcal{F}^{(s)}(a)-\frac{a^{2}}{4}\Box_{a}\mathcal{F}^{(s)}(a))*_{a}\delta h^{(s)}(a) .
\end{equation}
Zero order gauge invariance can be  checked easily by substitution of (\ref{4.5}) into this variation and use of the duality relation (\ref{4.15}) and identity (\ref{4.42}) taking into account tracelessness  of the gauge parameter (\ref{4.6}).
Now we turn to the generalization of Noether's procedure of the 2-2-4 case to the general s-s-2s interaction construction.
Noether's equation in this case looks like
\begin{equation}\label{4.58}
    \delta_{(1)}\mathcal{L}_{0}(h^{(s)}(a))+\delta_{0}\mathcal{L}_{1}(h^{(s)}(a),h^{(2s)}(b))=0 .
\end{equation}
And we would like to show that the solution of the latter is (with generalized Bell-Robinson current \cite{vanDam})
\begin{eqnarray}
  && \mathcal{L}_{1}(h^{(s)}(a),h^{(2s)}(b))=\frac{1}{2s}h^{(2s)}(z;b)*_{b}\Psi^{(2s)}_{(\Gamma)}(z;b) , \label{LL}\\
  && \Psi^{(2s)}_{(\Gamma)}(z;b)= \Gamma^{(s)}(b,a)*_{a}\Gamma^{(s)}(b,a)-\frac{b^{2}}{2(s+1)}\partial^{b}_{\mu}\Gamma^{(s)}(b,a)*_{a}\partial^{\mu}_{b}\Gamma^{(s)}(b,a) .\qquad\quad\label{LLL}
\end{eqnarray}

To prove this we must propose a first order variation of the spin s field with respect to a spin 2s gauge transformation. Remembering that Fronsdal's higher spin gauge potential is double traceless, we must
make sure that the same holds for the variation. Expanding the general variation in powers of $a^{2}$ \begin{equation}\label{4.46}
    \delta h^{(s)}(a)=\delta h^{(s)}_{(1)}(a)+a^{2}\delta h^{(s-2)}(a)+(a^{2})^{2}\delta h^{(s-4)}(a)+ \dots ,
\end{equation}
we see that the double tracelessness condition $\Box^{2}_{a} \delta h^{(s)}(a)=0$ expresses the third and
higher terms of the expansion (\ref{4.46}) through the first two free parameters $\delta h^{(s)}_{(1)}(a)$ and $\delta h^{(s-2)}(a)$\footnote{For completeness we present here the solution for $\delta h^{(s-4)}(a)$ following from the double tracelessness condition
\begin{eqnarray}
&& \delta h^{(s-4)}(a)=-\frac{1}{8\alpha_{1}\alpha_{2}} \left[\Box^{2}_{a}\delta h^{(s)}_{(1)}(a)+4\alpha_{1}\Box_{a}\delta h^{(s-2)}(a)\right] ,\nonumber\\
  && \alpha_{k}=D+2s-(4+2k) ,\quad k\in \{1,2\} . \nonumber
\end{eqnarray} }.
From the other hand Fronsdal's tensor is double traceless by definition and therefore all these $O(a^{4})$ terms are unimportant because they do not contribute to (\ref{4.45}). This leaves us freedom in the choice of $\delta h^{(s-2)}(a)$. Substituting (\ref{4.46}) in (\ref{4.45}) we discover that the following choice of  $\delta h^{(s-2)}(a)$
\begin{equation}\label{4.47}
    \delta h^{(s-2)}(a)=\frac{1}{2(D+2s-2)}\Box_{a}\delta h^{(s)}_{(1)}(a) ,
\end{equation}
reduces our variation (\ref{4.45}) to
\begin{equation}\label{4.48}
     \delta_{(1)}\mathcal{L}_{0}(h^{(s)}(a))=-\mathcal{F}^{(s)}(a)*_{a}\delta h^{(s)}_{(1)}(a) .
\end{equation}
Then we propose the following spin 2s transformation of the spin s potential
\begin{equation}\label{4.49}
    \delta h^{(s)}_{(1)}(a)= \tilde{\mathcal{U}}(b, a, 2, s)\epsilon^{2s-1}(z;b)*_{b}\Gamma^{(s)}(z;b,a) ,
\end{equation}
where
\begin{equation}\label{4.50}
    \tilde{\mathcal{U}}(b, a, 2, s)=\frac{(-1)^{s}}{(s-1)!}\prod^{s}_{k=2}\left[(\nabla\partial_{b})-\frac{1}{k}A_{b}(\nabla\partial_{a})\right] ,
\end{equation}
is operator dual to
\begin{equation}\label{4.51}
    [(b\nabla)-\frac{1}{2}(a\nabla)B_{a}]\mathcal{U}(b, a, 3, s)=\prod^{s}_{k=2}[(b\nabla)-\frac{1}{k}(a\nabla)B_{a}] ,
\end{equation}
with respect to the $*_{a,b}$ contraction product. Taking into account (\ref{4.35}) and Bianchi identities (\ref{4.41}) we get
 \begin{eqnarray}
   &&  \delta_{(1)}\mathcal{L}_{0}(h^{(s)}(a))=\epsilon^{2s-1}(z;b)*_{b}\Gamma^{(s)}(z;b,a)*_{a} [(b\nabla)-\frac{1}{2}(a\nabla)B_{a}]\mathcal{U}(b, a, 3, s)\mathcal{F}^{(s)}(z;a) \nonumber\\
   && = \epsilon^{2s-1}(z;b)*_{b}\Gamma^{(s)}(z;b,a)*_{a} \frac{1}{s(s-1)}[(b\nabla)-\frac{1}{2}(a\nabla)B_{a}]\Box_{b}\Gamma^{(s)}(z;b,a)\nonumber\\
   && = \epsilon^{2s-1}(z;b)*_{b}\Gamma^{(s)}(z;b,a)*_{a}\frac{1}{s}(\nabla\partial_{b})\Gamma^{(s)}(z;b,a)\nonumber\\
   && = -(b\nabla) \epsilon^{2s-1}(b)*_{b}\Gamma^{(s)}(b,a)*_{a}\Gamma^{(s)}(b,a)
   -\epsilon^{2s-1}(b)*_{b}\nabla_{\mu}\Gamma^{(s)}(b,a)*_{a}\frac{1}{s}\partial^{\mu}_{b}\Gamma^{(s)}(b,a) .\quad\quad\quad\quad
 \end{eqnarray}
Then using a secondary Bianchi identity  (\ref{4.40}) and a primary one (\ref{4.18}) one can show that
\begin{eqnarray}
  && -\epsilon^{2s-1}(b)*_{b}\nabla_{\mu}\Gamma^{(s)}(b,a)*_{a}\frac{1}{s}\partial^{\mu}_{b}\Gamma^{(s)}(b,a)\nonumber\\&&=
  \frac{1}{2s(s+1)(2s-1)}(\nabla\partial_{b})\epsilon^{2s-1}(b)*_{b}\partial^{b}_{\mu}\Gamma^{(s)}(b,a)*_{a}\partial^{\mu}_{b}\Gamma^{(s)}(b,a) .
\end{eqnarray}
Putting all together we see that the integrated first order interaction Lagrangian (\ref{LL})
supplemented with transformation (\ref{4.49}) for $h^{(s)}(a)$ and the standard zero order transformations for $h^{(2s)}(a)$
\begin{eqnarray}
  && \delta_{0} h^{(2s)}(z;b)=2s (b\nabla)\epsilon^{(2s-1)}(z;b) ,\\
  && \delta_{0} \Box_{b}h^{(2s)}(z;b)=4s (\nabla\partial_{b})\epsilon^{(2s-1)}(z;b) ,
\end{eqnarray}
completely solves Noether's equation (\ref{4.58}).
Note that here just as in the 2-2-4 case we did not obtain an interaction with lower spins because all derivatives included in the ansatz were used for the lifting to the second curvature.

\section{Conclusions}
We presented interaction Lagrangians for triplets of higher spin fields, a pair of which has equal spin $s_1$ whereas the third has even spin $s_2 \geq 2 s_1$. Besides the Lagrangians the next-to-leading order
of the gauge transformations is given. The fields of smaller spins appear combined into currents
of the Bell-Robinson form \cite{vanDam}. Remarkable is that for one such spin $s_2$ the interaction implies the existence of a whole ladder of interactions for smaller spins $s_2-2n \geq 2s_1$.

\subsection*{Acknowledgements}
This work is supported in part by Alexander von Humboldt Foundation under 3.4-Fokoop-ARM/1059429 and ANSEF 2009.
Work of K.M. was made with partial support of CRDF-NFSAT UCEP06/07.

\section*{Appendix}
\setcounter{equation}{0}
\renewcommand{\theequation}{A.\arabic{equation}}
\quad To manipulate reshuffling of different sets of indices we employ two differentials with respect to $a$ and $b$, e.g.
\begin{eqnarray}
A_{b} = (a\partial_{b}) , \\
\label{2.11}
B_{a} = (b\partial_{a}) .
\label{2.12}
\end{eqnarray}
Then we see that operators $A_{b}, a^{2}, b^{2}$ are dual (or adjoint) to $B_{a},\Box_{a},\Box_{b}$ with respect to the "star" product of tensors with two sets of symmetrized indices  (\ref{4.12})
\begin{eqnarray}
    \frac{1}{n}A_{b}f^{(m-1,n)}(a,b)*_{a,b} g^{(m,n-1)}(a,b)&=& f^{(m-1,n)}(a,b)*_{a,b} \frac{1}{m}B_{a}g^{(m,n-1)}(a,b) ,\label{4.13}\\
    a^{2}f^{(m-2,n)}(a,b)*_{a,b} g^{(m,n)}(a,b)&=&f^{(m-2,n)}(a,b)*_{a,b} \frac{1}{m(m-1)}\Box_{a} g^{(m,n)}(a,b) . \nonumber\\\label{4.14}
\end{eqnarray}
In the same fashion gradients and divergences are dual with respect to the full scalar product in the space $(z,a,b)$
\begin{eqnarray}
  (a\nabla)f^{(m-1,n)}(z;a,b)*_{a,b} g^{(m,n)}(z;a,b) &=& -f^{(m-1,n)}(z;a,b)*_{a,b}\frac{1}{m}(\nabla\partial_{a}) g^{(m,n)}(z;a,b) .\nonumber\\ \label{4.15}
  \end{eqnarray}
Analogous equations can be formulated for the operators $b^{2}$ or $b\nabla$.

Now one can prove that \cite{DF, MR5}:
\begin{equation}
A_{b}\Gamma^{(s)}(z;a,b) = B_{a}\Gamma^{(s)}(z;a,b) = 0 .
\label{4.18}
\end{equation}
These "primary Bianchi identities" are manifestations of the hidden antisymmetry.
The n-th deWit-Freedman-Cristoffel symbol is
\begin{eqnarray}
\Gamma_{(n)}^{(s)}(z;b,a)&&\equiv \Gamma^{(s)}_{(n)\rho_{1}...\rho_{n},\mu_{1}...\mu_{\ell}}b^{\rho_{1}}...b^{\rho_{n}}
a^{\mu_{1}}...a^{\mu_{\ell}}\nonumber\\&&=[(b\nabla)-\frac{1}{n}(a\nabla)B_{a}]\Gamma_{(n-1)}^{(s)}(z;b,a) ,
\end{eqnarray}
or in another way
\begin{equation}
\Gamma_{(n)}^{(s)}(z;b,a)=(\prod_{k=1}^{s}[(b\nabla)-\frac{1}{k}(a\nabla)B_{a}])h^{(s)}(z;a) .
\end{equation}
Using the following commutation relations
\begin{eqnarray}
&&[B_{a},(a\nabla)]=(b\nabla),\label{4.21}\\
&&[B_{a}^{k},(a\nabla)]=kB_{a}^{k-1}(b\nabla),\\
&&[B_{a},(a\nabla)^{k}]=k(b\nabla)(a\nabla)^{k-1},\\
&&\Box_{b}(b\nabla)^{i}=i(i-1)(b\nabla)^{i-2}\Box,\\
&&\partial^{b}_{\mu}(b\nabla)^{i}\partial_{b}^{\mu}B_{a}^{j}=ij(b\nabla)^{i-1}B_{a}^{j-1}(\nabla\partial_{a}),\\
&&\Box_{b}B_{a}^{j}=j(j-1)B_{a}^{j-2}\Box_{a} ,\label{4.26}
\end{eqnarray}
and mathematical induction we can prove that
\begin{eqnarray}
\Gamma_{(n)}^{(s)}(z;b,a)=\sum_{k=0}^{n}\frac{(-1)^{k}}{k!}(b\nabla)^{n-k}(a\nabla)^{k}B_{a}^{k}h^{(s)}(z;a) .\label{gumar}
\end{eqnarray}
The gauge variation of the n-th Cristoffel symbol is
\begin{eqnarray}
&&\delta \Gamma_{(n)}^{(s)}(z;b,a)=\frac{(-1)^{n}}{n!}(a\nabla)^{n+1}B_{a}^{n}\epsilon^{(s-1)}(z;a) ,
\end{eqnarray}
putting here $n=s$ we obtain gauge invariance for the curvature
\begin{equation}
\delta \Gamma_{(s)}^{(s)}(z;b,a)=0 .
\end{equation}
Tracelessness of the gauge parameter (\ref{4.6})
implies that b-traces of all Cristoffel symbols are gauge invariant
\begin{eqnarray}
&& \Box_{b}\delta \Gamma_{(n)}^{(s)}(z;b,a)=\frac{(-1)^{n}}{(n-2)!}(a\nabla)^{n+1}B_{a}^{n-2}\Box_{a}\epsilon^{(s-1)}(z;a)=0 .
\end{eqnarray}
Thus for the second order gauge invariant field equation we can use the trace of the second Cristoffel symbol,
the so called Fronsdal tensor:
\begin{eqnarray}
\mathcal{F}^{(s)}(z;a)&=&\frac{1}{2}\Box_{b}\Gamma_{(2)}^{(s)}(z;b,a)\nonumber\\
&=&\Box h^{(s)}(z;a)-(a\nabla)(\nabla\partial_{a})h^{(s)}(z;a)
+\frac{1}{2}(a\nabla)^{2}\Box_{a}h^{(s)}(z;a) .\quad\label{4.32}
\end{eqnarray}
Using equation (\ref{gumar}) for Cristoffel symbols and
after long calculations we obtain the following expression
\begin{eqnarray}
&&\Box_{b}\Gamma_{(n)}^{(s)}(z;b,a)\nonumber\\&&=\sum_{k=0}^{n-2}\frac{(-1)^{k}}{k!}(n-k)(n-k-1)
(b\nabla)^{n-k-2}(a\nabla)^{k}B_{a}^{k}\mathcal{F}^{(s)}(z;a) .
\end{eqnarray}
We have expressed the b-trace of any $\Gamma_{(n)}^{(s)}$ through the Fronsdal tensor or the b-trace of the second Cristoffel symbol, but this is not the whole story. Using mathematical induction and (\ref{4.21})-(\ref{4.26}) again we can show that
\begin{eqnarray}
&&\sum_{k=0}^{n-2}\frac{(-1)^{k}}{k!}(n-k)(n-k-1)(b\nabla)^{n-k-2}(a\nabla)^{k}B_{a}^{k}\mathcal{F}^{(s)}(z;a)\nonumber\\
&&\quad\quad\quad\quad=n(n-1)(\prod^{n}_{k=3}[(b\nabla)-\frac{1}{k}(a\nabla)B_{a}])\mathcal{F}^{(s)}(z;a) .
\end{eqnarray}
In particular for the trace of the curvature we can write
\begin{eqnarray}
&&\Box_{b}\Gamma^{(s)}(z;b,a)=s(s-1)\mathcal{U}(a,b,3,s)\mathcal{F}^{(s)}(z;a) ,\label{4.35}
\end{eqnarray}
where we introduced an operator mapping the Fronsdal tensor on the trace of the curvature
\begin{equation}\label{4.36}
   \mathcal{U}(a,b,3,s)=\prod^{s}_{k=3}[(b\nabla)-\frac{1}{k}(a\nabla)B_{a}] .
\end{equation}
Now let us consider this curvature in more detail. First we have the symmetry under exchange of $a$ and $b$
\begin{equation}
\Gamma^{(s)}(z;a,b) = \Gamma^{(s)}(z;b,a) .
\end{equation}
Therefore the operation "$a$-trace" can be defined by (\ref{4.35}) with exchange of $a$ and $b$ at the end.
The mixed trace of the curvature can be expressed through the $a$ or $b$ traces using "primary Bianchi identities" (\ref{4.18})
\begin{equation}\label{4.38}
(\partial_{a}\partial_{b})\Gamma^{(s)}(z;b,a)=-\frac{1}{2}B_{a}\Box_{b}\Gamma^{(s)}(z;b,a)=
-\frac{1}{2}A_{b}\Box_{a}\Gamma^{(s)}(z;b,a) .
\end{equation}

The next interesting properties of the higher spin curvature and corresponding Ricci tensors are so called generalized secondary or differential Bianchi identities. We can formulate  these identities  in our notation in the following compressed form ($[\dots]$ denotes antisymmetrization )
\begin{equation}\label{4.39}
\frac{\partial}{\partial a^{[\mu}}\frac{\partial}{\partial b^{\nu}}\nabla_{\lambda]}\Gamma^{(s)}(z;a,b)= 0 .
\end{equation}
This relation can be checked directly from representation (\ref{gumar}). Then contracting with $a^{\mu}$ and $b^{\nu}$ we get a symmetrized form of (\ref{4.39})
\begin{equation}\label{4.40}
    s \nabla_{\mu}\Gamma^{(s)}(z;a,b)=(a\nabla)\partial^{a}_{\mu}\Gamma^{(s)}(z;a,b)+(b\nabla)\partial^{b}_{\mu}\Gamma^{(s)}(z;a,b) .
\end{equation}
Now we can contract (\ref{4.40}) with a $\partial^{\mu}_{b}$ and using (\ref{4.38}) obtain a connection between the divergence and the trace of the curvature
\begin{equation}\label{4.41}
    (s-1)(\nabla\partial_{b})\Gamma^{(s)}(z;a,b)=[(b\nabla)-\frac{1}{2}(a\nabla)B_{a}]\Box_{b}\Gamma^{(s)}(z;a,b) .
\end{equation}
These two identities with a similar identity for the Fronsdal tensor
\begin{equation}\label{4.42}
    (\nabla\partial_{a})\mathcal{F}^{(s)}(z;a)=\frac{1}{2}(a\nabla)\Box_{a}\mathcal{F}^{(s)}(z;a) ,
\end{equation}
play an important role for the construction of the interaction Lagrangian.


\begin{thebibliography}{100}
\bibitem{MMR}
R.~Manvelyan and K.~Mkrtchyan,
``Conformal invariant interaction of a scalar field with the higher spin
  field in $AdS_{D}$,'' [arXiv:0903.0058 [hep-th]].
\bibitem{MR}
R.~Manvelyan and W.~R\"uhl, ``Conformal coupling of higher spin
gauge fields to a scalar field in AdS(4) and generalized Weyl
invariance,'' Phys.\ Lett.\ B {\bf 593} (2004) 253,
[arXiv:hep-th/0403241].
\bibitem{vanDam}
  F.~A.~Berends, G.~J.~H.~Burgers and H.~van Dam,
  ``Explicit Construction Of Conserved Currents For Massless Fields Of
  Arbitrary Spin,'' Nucl.\ Phys.\  B {\bf 271} (1986) 429;
   F.~A.~Berends, G.~J.~H.~Burgers and H.~Van Dam,
  ``On Spin Three Selfinteractions,''
  Z.\ Phys.\  C {\bf 24} (1984) 247;
  F.~A.~Berends, G.~J.~H.~Burgers and H.~van Dam,
  ``On The Theoretical Problems In Constructing Interactions Involving Higher
  Spin Massless Particles,''
  Nucl.\ Phys.\  B {\bf 260} (1985) 295.
\bibitem{Vasiliev}
  E.~S.~Fradkin and M.~A.~Vasiliev, ``On The Gravitational Interaction
  Of Massless Higher Spin Fields,'' Phys.\ Lett.\ B {\bf 189} (1987)
  89.
   E.~S.~Fradkin and M.~A.~Vasiliev, ``Cubic Interaction In
  Extended Theories Of Massless Higher Spin Fields,'' Nucl.\ Phys.\ B
  {\bf 291} (1987) 141.
\bibitem{Damour}
  T.~Damour and S.~Deser,
  ``Geometry of spin 3 gauge theories,''
  Annales Poincare Phys.\ Theor.\  {\bf 47}, 277 (1987);
 T.~Damour and S.~Deser,
  ``Higher derivative interactions of higher spin gauge fields,''
  Class.\ Quant.\ Grav.\  {\bf 4}, L95 (1987).
\bibitem{review}
M. A. Vasiliev, ``Higher Spin Gauge Theories in Various Dimensions'',
Fortsch. Phys. 52, 702 (2004) [arXiv:hep-th/0401177].
 X. Bekaert, S. Cnockaert, C. Iazeolla and M. A. Vasiliev,
 ``Nonlinear higher spin theories in various dimensions
'', [arXiv:hep-th/0503128].
 D. Sorokin,``Introduction to the Classical Theory of Higher Spins'' AIP Conf. Proc. 767, 172
(2005); [arXiv:hep-th/0405069]. N. Bouatta, G. Compere and A. Sagnotti, ``An Introduction to Free Higher-Spin Fields''; [arXiv:hep-th/0409068].
\bibitem{Metsaev}
  R.~R.~Metsaev,
  ``Cubic interaction vertices for massive and massless higher spin fields,''
  Nucl.\ Phys.\  B {\bf 759} (2006) 147
  [arXiv:hep-th/0512342];R.~R.~Metsaev,
  ``Cubic interaction vertices for fermionic and bosonic arbitrary spin
  fields,''
  arXiv:0712.3526 [hep-th].
\bibitem{ouvry}
  I.~G.~Koh, S.~Ouvry, ``Interacting gauge fields of any spin and symmetry,''
  Phys. Lett. B {\bf 179} (1986) 115; Erratum-ibid. {\bf 183} B (1987) 434.
\bibitem{boulanger}
  Nicolas Boulanger, Serge Leclercq, Per Sundell,
  ``On The Uniqueness of Minimal Coupling in Higher-Spin Gauge Theory,''
  JHEP 0808:056,2008;  [arXiv:0805.2764 [hep-th]].
  Xavier Bekaert, Nicolas Boulanger, Sandrine Cnockaert, Serge Leclercq,
  ``On killing tensors and cubic vertices in higher-spin gauge theories,''
  Fortsch. Phys. {\bf 54} (2006) 282-290; [arXiv:hep-th/0602092].
\bibitem{Petkou}
  A.~Fotopoulos, N.~Irges, A.~C.~Petkou and M.~Tsulaia,
  ``Higher-Spin Gauge Fields Interacting with Scalars: The Lagrangian Cubic
  Vertex,''
  JHEP {\bf 0710} (2007) 021;
  [arXiv:0708.1399 [hep-th]].
  I.~L.~Buchbinder, A.~Fotopoulos, A.~C.~Petkou and M.~Tsulaia,
  ``Constructing the cubic interaction vertex of higher spin gauge fields,''
  Phys.\ Rev.\  D {\bf 74} (2006) 105018;
  [arXiv:hep-th/0609082].
\bibitem{Kleb}
  I.~R.~Klebanov and A.~M.~Polyakov, ``AdS dual of the critical O(N)
  vector model,'' Phys.\ Lett.\ B {\bf 550} (2002) 213;
  [arXiv:hep-th/0210114].
\bibitem{MMR1}
R.~Manvelyan, K.~Mkrtchyan and W.~R\"uhl,
  ``Ultraviolet behaviour of higher spin gauge field propagators and one loop
  mass renormalization,''
  Nucl.\ Phys.\  B {\bf 803} (2008) 405
  [arXiv:0804.1211 [hep-th]].
\bibitem{Ruehl}
  W.~R\"uhl, ``The masses of gauge fields in higher spin field theory on
  AdS(4),'' Phys.Lett. B {\bf 605} (2005) 413; [arXiv:hep-th/0409252]; the results
  presented here are based on extensive calculations performed by K. Lang and
  W.~R\"uhl, Nucl. Phys. B {\bf 400} (1993) 597.
\bibitem{MR1}
  R.~Manvelyan and W.~R\"uhl,
  ``The off-shell behaviour of propagators and the Goldstone field in  higher
  spin gauge theory on AdS(d+1) space,''
  Nucl.\ Phys.\  B {\bf 717} (2005) 3;
  [arXiv:hep-th/0502123].
 \bibitem{MR2}
 R.~Manvelyan and W.~R\"uhl,
  ``The masses of gauge fields in higher spin field theory on the bulk of
  AdS(4),''
  Phys.\ Lett.\  B {\bf 613} (2005) 197;
  [arXiv:hep-th/0412252].
 \bibitem{MR3}
  R.~Manvelyan and W.~R\"uhl,
  ``The structure of the trace anomaly of higher spin conformal currents in the
  bulk of AdS(4),''
  Nucl.\ Phys.\  B {\bf 751}, (2006) 285;
  [arXiv:hep-th/0602067].
\bibitem{MR4} R.~Manvelyan and W.~R\"uhl,
  ``The quantum one loop trace anomaly of the higher spin conformal  conserved
  currents in the bulk of AdS(4),''
  Nucl.\ Phys.\  B {\bf 733} (2006) 104;
  [arXiv:hep-th/0506185].
\bibitem{MR5}
  R.~Manvelyan and W.~R\"uhl,
  ``Generalized Curvature and Ricci Tensors for a Higher Spin Potential and the
  Trace Anomaly in External Higher Spin Fields in $AdS_{4}$ Space,''
  Nucl.\ Phys.\  B {\bf 796} (2008) 457;
  [arXiv:0710.0952 [hep-th]].
\bibitem{Frons}
C.~Fronsdal,
 ``Singletons And Massless, Integral Spin Fields On De Sitter Space (Elementary
Particles In A Curved Space Vii),'' Phys.\ Rev.\ D {\bf 20},
(1979) 848;``Massless Fields With Integer Spin,'' Phys.\ Rev.\ D {\bf
18} (1978) 3624.
\bibitem{DF}
B.~deWit and D.Z.~Freedman, ``Systematics of higher spin gauge
fields,'' Phys. Review D \textbf{21} (1980), 358-367.
\bibitem{MR6}
  R.~Manvelyan and W.~R\"uhl,
  ``The Generalized Curvature and Christoffel Symbols for a Higher Spin
  Potential in $AdS_{d+1}$ Space,''
  Nucl.\ Phys.\  B {\bf 797}, 371 (2008)
  [arXiv:0705.3528 [hep-th]].
\bibitem{PF}
  M.~Fierz and W.~Pauli,
  ``On relativistic wave equations for particles of arbitrary spin in an
  electromagnetic field,''
  Proc.\ Roy.\ Soc.\ Lond.\  A {\bf 173} (1939) 211.
\end{thebibliography}
\end{document}